\documentstyle[12pt]{article}

\newcommand{\disregard}[1]{}

\newcommand{\be}{\begin{equation}}
\newcommand{\ee}{\end{equation}}
\newcommand{\ba}{\begin{array}}
\newcommand{\ea}{\end{array}}
\newcommand{\bt}{\begin{tabular}}
\newcommand{\et}{\end{tabular}}
\newcommand{\bc}{\begin{center}}
\newcommand{\ec}{\end{center}}
\newcommand{\bn}{\begin{eqnarray}}
\newcommand{\en}{\end{eqnarray}}
\newcommand{\spc}{{\ }}
\newcommand{\citeq}[1]{Eq.~(\ref{#1})}
\newcommand{\citeqs}[1]{Eqs.~(\ref{#1})}
\newcommand{\citefig}[1]{Fig.~\ref{#1}}
\newcommand{\citefigs}[1]{Figs.~\ref{#1}}
\newcommand{\citesec}[1]{Section~\ref{#1}}
\newcommand{\citetab}[1]{Table~\ref{#1}}
\newcommand{\citetabs}[1]{Tables~\ref{#1}}

\newcommand{\deltan}       {\delta_{\rm N}}
\newcommand{\deltap}       {\delta_{\rm P}}
\newcommand{\tDeltan}{\tilde\Delta_{\rm N}}
\newcommand{\tDeltap}{\tilde\Delta_{\rm P}}
\newcommand{\shellgap}{shell-gap}
\newcommand{\bbox}[1]{\mbox{\boldmath{$#1$}}}
\title      {
              Closed shells at drip-line nuclei}
\author  { J. Dobaczewski,$^a$
           W. Nazarewicz,$^{a,b,c,d}$
           and T.R. Werner$^{a,b,c}$\\
          $^a$Institute of Theoretical
                Physics, Warsaw University\\
                ul. Ho\.za 69, 00-681 Warsaw, Poland\\
           $^b$Physics Division,  Oak Ridge National Laboratory\\
                P.O. Box 2008, Oak Ridge, Tennessee 37831, USA\\
           $^c$Department of Physics and Astronomy,
                University of Tennessee\\
                Knoxville, Tennessee 37996, USA\\
           $^d$Joint Institute for Heavy-Ion Research, Oak Ridge\\
                Tennessee 37831, USA
         }
\date{{}}

\begin{document}
\maketitle

%\noi
%PACS: 21.10.Dr, 21.10.Pc, 21.30+y, 21.60.Jz

%\preprint{IFT/20/94}

\begin{abstract}
The shell structure of magic nuclei far from stability
is discussed in terms of
the self-consistent spherical Hartree-Fock-Bogoliubov theory.
In particular, the sensitivity of the {\shellgap} sizes
and the two-neutron separation energies  to the choice of
particle-hole and particle-particle components of
the effective interaction is investigated.
\end{abstract}

\section{Introduction}
\label{sec1}

Shell effects in nuclei lead to characteristic irregularities in
otherwise smooth behaviour of global nuclear observables when
they are analyzed as functions of $Z$, $N$, and $A$.
This applies in particular to
nuclear masses, or nuclear binding energies, which
conspicuously show that the closed-shell nuclei
are bound stronger than open-shell systems
%\cite{[Boh69],[Mye82],[Rag84]}.
%\citemany{[Boh69],[Rag84]}.
[1-3].
 Rather simple
one-body potentials can account for the values of particle
numbers where strong shell effects occur
 [magic numbers, equal to 8, 20, 28, 40, 50, 82, and
(for neutrons only)
126].

Through the binding energies, many other nuclear structure
properties depend on the strength of shell effects
(for discussion of various
binding-energy relations, see Ref.{\spc}~\cite{[Jen84]}, Sec. 2).
 This is
particularly true for the particle-emission lifetimes and/or
particle-capture probabilities which dramatically depend on
accessible energies.  In stable nuclei these dependencies are
well studied and analyzed. However, for nuclei far from stability,
the relevant experimental data are scarce or simply nonexistant.
 For example, the
nucleosynthesis of heavy elements in supernova explosions
depends crucially on the values of beta decay rates and the neutron
capture cross sections in very neutron-rich nuclei which do not
exist in terrestrial conditions \cite{[How93],[Kra93]}.
 A prediction of nuclear shell
effects far from stability, based on the experience gained by
studying stable nuclei, is an important and difficult  challenge to
nuclear structure theory.

In the present contribution, we aim at quantitatively describing
the magnitude of shell effects in nuclei close to the neutron drip
line. The description is based
on the self-consistent
Hartree-Fock-Bogoliubov (HFB) theory
with
effective interactions   fitted to known nuclei.
In the particle-hole (p-h) channel, we employ
the Skyrme interaction.
In the particle-particle (p-p, pairing) channel, several contact forces are
used.

All details of the
method follow Ref.{\spc}\cite{[Dob84]},
where the HFB theory in spatial coordinates has been introduced,
and the Skyrme parametrization SkP has been obtained by a fit to
properties of
several magic nuclei and to the chain of the tin isotopes. In the
present study, we
also use the HFB method with other standard Skyrme-force
parametrizations such as SIII \cite{[Bei75]} and SkM* \cite{[Bar82]}.

In drip-line nuclei, special attention should be paid to the
description of pairing correlations.  In particular, the condition
\cite{[Bei75a],[Smo93]}:
\be\label{sep}
S_n \approx -\lambda_n - \Delta =0,
\ee
which requires that at the drip line the neutron separation
energy, $S_n$, vanishes, shows that pairing effects are crucial for
determining the position of the one-neutron drip line.
{}From \citeq{sep} it follows
that properties of the mean field, characterized by
the Fermi energy, $\lambda$,
(determined by the p-h component of the effective
interaction) and those of the pairing field, $\Delta$, (determined by
the p-p part of the effective interaction)
are equally important far from the beta-stability line.
Consequently, the choice of the pairing interaction
is  vital for analyzing shell effects
in drip-line nuclei.

In the present study,   results obtained with several
p-h and pairing interactions (see
\citesec{sec2}) are confronted with the data.
A comparison of theoretical predictions for the
two-neutron separation energies, $S_{2n}$, with experimental values
is given in \citesec{sec3}, while in \citesec{sec4}
the sizes  of calculated neutron shell gaps for very
neutron-rich nuclei are discussed.

\section{Pairing interaction}
\label{sec2}

\subsection{Magnitude of pairing correlations}
\label{sec2a}

Odd-even staggering of nuclear masses, i.e., the fact that
nuclides with odd numbers of protons or neutrons are found to have
smaller binding energy than the
arithmetic mean of the binding energies of the two neighboring even-even
nuclei,
 is a direct consequence of nuclear
pairing correlations.  The strength of pairing
interaction is often adjusted so as to reproduce
the odd-even mass difference
\be\label{oddeven}
\deltan = \frac{1}{2}\bigl(B(N-1,Z)+B(N+1,Z)\bigr) - B(N,Z).
\ee
In particular, in Ref.{\spc}\cite{[Dob84]},
a certain linear combination of the parameters of
the Skyrme force
has been adjusted
to fit the values of the odd-even mass difference in
the tin isotopes. The resulting
Skyrme interaction, SkP, describes
both p-h and p-p channels.

In \citetab{tab2} we present the neutron odd-even mass difference
for proton-magic nuclei  obtained
in the HFB+SkP approach.  For every
magic proton number, calculations were performed for one odd
isotope near the bottom of the stability valley and for its two
even-even neighbours. In an  odd isotope, a
particle state was blocked (by
the method given in Ref.{\spc}\cite{[Dob84]}), and then the lowest energy
was selected among different configurations.
The experimental values of
$\deltan^{\rm exp}$, determined from experimental binding
energies of Ref.{\spc}\cite{[Aud93]}, are also given in \citetab{tab2}. In
\citetab{tab3} we present the proton odd-even mass differences
with
all definitions and methods {\em mutatis mutandis} analogous to
those used for neutrons.

In \citetabs{tab2} and \ref{tab3} we also present
the values of experimental and calculated pairing gaps.
The former are derived from finite-difference formulas
involving experimental binding energies \cite{[Mol92b]},
and the latter are the average values
defined according to Ref.{\spc}\cite{[Dob84]}.
As discussed in Ref.{\spc}\cite{[Boh69]}, the experimental pairing gaps
are, in general, larger than the odd-even mass staggering
(\ref{oddeven}).

\citetabs{tab2} and \ref{tab3} show that the theoretical
values of $\deltan$ and $\deltap$ are generally much
smaller than the experimental values. The only exception is the
odd-even mass difference
for $^{119}$Sn which was actually used when fitting the
SkP force.
Based on this result,  one may be tempted to
refit the force parameters to increase the average magnitude of
pairing correlations and improve the agreement between experimental and
theoretical values of $\delta$.
However, a comparison of pairing-type
two-body matrix elements of the SkP force with experimentally-deduced
pairing matrix elements \cite{[Sch76]}
indicates that they have, on the average, correct
magnitude.  We are therefore led to the conclusion  that it is
the blocking
mechanism that is responsible for too low values of
$\deltan$.

In fact,
the strength of pairing interaction in nuclei is never far
away from the critical value at which, in the mean field
approximation, the static  pairing correlations vanish
\cite{[RS80]}.  This situation
precludes the use of
the blocking approximation, because
 the critical value of pairing strength
is rather large in many-quasiparticle configurations.
 For odd nuclei,
particle number projection should, in principle, be used to describe
pairing correlations near the phase transition
region \cite{[Man65]}.

On the other hand, a comparison of experimental and theoretical
(average) pairing gaps is less conclusive than a direct comparison
of nuclear masses entering the odd-even staggering (\ref{oddeven}).
This is so, because the staggering should specifically depend on the last
occupied orbit near the Fermi energy, while the average pairing gap
is a global property depending on all single-particle states.
Until a better theory of pairing correlations in odd nuclei is applied,
we do not have the correct tools which would allow
the incorporation of experimental information
on the odd-even staggering or on the experimental pairing gaps in
a {\em detailed} adjustment of forces in the p-p channel.
Therefore, in the present
study, we consider even-mass nuclei only, and consequently, we use average
pairing gaps, $\tDeltan$ and $\tDeltap$, as {\em qualitative measures} of the
pairing correlations.

The last columns of \citetabs{tab2} and \ref{tab3} display
proton and neutron average pairing gaps at proton and neutron magic
numbers, respectively. Neither $N$=40 nor $Z$=40 shell gaps in
this calculations are large enough to destroy pairing correlations.
Static pairing is also obtained for the
$N$=20 and 28 systems.
This suggests that the
magnitude of
pairing correlations predicted by SkP  is slightly too
large in light nuclei.
However,  this is a rather subtle discrepancy, because at particle
numbers 20 and 28 (both
for protons and neutrons)
the total energies of paired and unpaired solutions are very close. For
the protons, the unpaired solution is slightly lower in energy while
for the neutrons, the paired solution is favoured.

\subsection{Pairing interactions}
\label{sec2b}

In the present paper we use three kinds of pairing forces.
{}Firstly, we present the HFB results with  the Skyrme
force SkP.
As discussed in Refs.{\spc}%
%\cite{[War79],[Gia81],[War83],[Dob84]}
%\citemany{[War79],[War83]}\cite{[Dob84]}
[7,17-19]
and in the previous Section, the
Skyrme parametrization can be adjusted in such a
way that for states near the Fermi
energy it has
attractive pairing matrix elements. The SkP interaction fulfills this
requirement.
Secondly, we employ a simple contact interaction
%\cite{[Ton79],[Zver85],[Kri90]}:
%\citemany{[Ton79],[Kri90]}:
[20-22]:
\be\label{e1}
V^\delta(\bbox{r},\bbox{r}') = V_0\delta(\bbox{r}-\bbox{r}'),
\ee
and,  thirdly, the contact interaction depending on the total
(isoscalar) density
of particles $\rho(\bbox{r})$
%\cite{[Cha76],[Sta92],[Taj93b],[Fay94]},
%\citemany{[Cha76],[Fay94]},
[23-26],
\be\label{e2}
V^{\delta\rho}(\bbox{r},\bbox{r}')
     = \left(V_0+V_3\rho^\gamma\right)\delta(\bbox{r}-\bbox{r}').
\ee
Since the Skyrme force is also a contact interaction, the use of
the SkP parametrization in the p-p channel amounts to adding
momentum-dependent and spin-exchange terms to simple forces given in
\citeqs{e1} and (\ref{e2}).

The form of the density dependence used in \citeq{e2} is
completely analogous to the standard Skyrme force, and hence we
use an analogous notation.  Parameters $V_0$, $V_3$, and
$\gamma$ should be chosen in such a way that the interaction is
strongly attractive outside the nucleus (i.e., $V_0$ should
be  large and
negative) and weak inside.  In this way, the
pair density becomes concentrated at the
surface region while for the pure contact force, \citeq{e1}, it is
spread throughout the nuclear interior
%\cite{[Sta92],[Bel87],[Dob94a]}.
%\citemany{[Sta92],[Dob94a]}.
[24-28].

Parameters of phenomenological pairing interactions
depend
 on the features of the mean field in the p-h channel, and
especially on the density of single-particle states near the
{}Fermi level. Therefore, for every parametrization of the
Skyrme force used in the p-h channel, one should use different
values of parameters defining pairing  interaction.
 In the present work,  we have adjusted the strength,
$V_0$, of the contact force (\ref{e1}) in the following way:  For every
version of the Skyrme force used in the p-h channel,
spherical HFB calculations (without blocking)
were performed for $^{119}$Sn
(see \citesec{sec2a}).  Then the
values of $V_0$ were fixed by requiring that in each case the
average pairing gap, $\tDeltan$, was  equal to 1.256~MeV which is
the value obtained in the HFB+SkP calculations (see \citetab{tab2}).
 When the contact interaction (\ref{e1}) is
used in the p-p channel and the SIII, SkM*, and SkP
parametrizations are used in the p-h channel, we denote the
resulting combined forces by SIII$^\delta$, SkM$^\delta$, and
SkP$^\delta$, respectively.

Since the density dependence in \citeq{e2} introduces two
additional parameters, for the purpose of the present study, we
have decided to use  the same values of $V_0$ and $\gamma$ as those used
in the p-h channel, and then to fit $V_3$ so as to reproduce
the  value of $\tDeltan$ in $^{119}$Sn. These
forces are denoted by SIII$^{\delta\rho}$,
SkM$^{\delta\rho}$, and SkP$^{\delta\rho}$, respectively.  Parameters of
pairing forces used in the present study are listed
in \citetab{tab1}.  In all cases, the phase space used, when
solving the HFB equations, was chosen according to the prescription
given in Ref.{\spc}\cite{[Dob84]}.

Average pairing gaps for semi-magic nuclei listed in
\citetabs{tab2} and \ref{tab3} are plotted in \citefigs{fig1}
and \ref{fig2} for neutrons and protons, respectively.
Open and full symbols refer to pairing forces (\ref{e1}) and (\ref{e2}),
respectively. The results obtained
with the SkP force agree quite
satisfactorily with experimental values of
odd-even staggering, apart from the neutrons  in
$^{59}_{31}$Ni$_{28}$ where all the forces considered fail.
 Neutron gaps are,
in general, reproduced better than the proton gaps which
for some forces are
often much too small.
Therefore, for protons one sometimes has to use a
larger interaction
strength $V_0$ than for neutrons \cite{[Kri90],[Taj93b]}.
Here we are mostly interested in neutron properties and therefore we
use a common value for both types of particles. Moreover, the SkP
parametrization does not make such a distinction and, nevertheless,
it provides a correct description of neutron and proton pairing
properties simultaneously.

Density-dependent pairing interactions
work better
in light nuclei. On the other hand,
the pure contact force seems to work better
for heavier systems.   No general pattern appears as far
as differences between various forces used in the p-h channel are
concerned. Most likely, these differences are related
to details of the placement of single-particle levels near the Fermi
energy which significantly vary from force to force.

As expressed by \citeq{sep},
to obtain a reasonable description of drip-line nuclei,
 pairing correlations have
to be considered. However,
compared with many detailed studies of Skyrme forces in the p-h channel,
much less effort has been devoted  to adjusting
properties of pairing interactions.
 In stable nuclei,
 pairing correlations can be incorporated
to some extent by using the BCS approximation, but
when departing from the beta-stability valley,
use of the HFB method is a must \cite{[Dob84],[Naz94a]}.
 A simple one-parameter fit, carried out in
the present study,  makes it possible to reproduce an overall magnitude
of pairing correlations in the few cases discussed.
The forces listed in \citetab{tab1} should be viewed as zero-order
approximations, enabling us to perform the HFB analysis with
standard Skyrme interactions.
Certainly,  more extensive fits of contact forces
(possibly with additional terms included) are necessary to
obtain better overall agreement with experiment. In this respect,
the relative
success of the pairing SkP force  may serve as an
indication that such an  improvement is indeed possible.

\section{Two-neutron separation energies}
\label{sec3}

Experimental two-neutron separation energies \cite{[Aud93]},
$S_{2n}$=$B(N,Z)$--$B(N-2,Z)$, are plotted in \citefig{fig3}(a)
for five chains of proton-magic isotopes. Values quoted
in Ref.{\spc}\cite{[Aud93]} as ``from systematics'' are denoted by
open symbols. Below the $S_{2n}$$\simeq$10~MeV line, experimental data
are not available, and  dramatic extrapolations in $N$
are required to reach the
region of the r-process path ($S_{2n}$$\approx$2 MeV) or the
two-neutron drip line
($S_{2n}$=0).

Among proton-magic, medium-heavy and heavy nuclei,
there exists information
about six closed neutron shells.
Namely, the experimental data are available for
two calcium isotopes at $N$=20 and
$N$=28, two nickel isotopes at $N$=28 and $N$=40, and the
zirconium and lead isotopes with  $N$=50 and $N$=126, respectively.
In all these cases, the values of $S_{2n}$ exhibit sudden
jumps of varying magnitude \cite{[Aud93]}.
The size of the shell gap can
be directly related to a difference between  two-neutron
separation energies,
\be\label{e3}
S_{2n}^N - S_{2n}^{N+2} = -B(N-2,Z) + 2B(N,Z) - B(N+2,Z),
\ee
i.e., is related to the second derivative of the binding energy.
Hence, in order to establish the size of the neutron shell gap,
the mass of the doubly magic nucleus and the masses of
its two neighbouring even isotopes have to be known. Similar information
can be obtained by analyzing quantities $\Delta_{2n}$ introduced
in \cite{[Jen84]}. However, they involve masses of four
isotopes and, therefore, they are
less convenient. It should be stressed that in our study
we do not aim at subtracting
the smooth background from the experimental masses, but rather use
indicators (\ref{e3}) to compare theoretical and experimental masses.

A very long chain of known tin isotopes starts and terminates
around
neutron magic numbers $N$=50 and $N$=82.
Unfortunately, so far, the masses of $^{100}$Sn and $^{134}$Sn
have not been measured, and $^{98}$Sn is probably proton-unstable.
Therefore,  for the purpose of the present
study, we establish
 the size of the $N$=82 shell gap
at $Z$=50 by
taking
the mass of $^{134}$Sn
 from the systematic trend proposed
in Ref.{\spc}\cite{[Aud93]}.

The neutron subshell effect at $N$=40 is rather weak, and the experimental
$S_{2n}$-values for the nickel
isotopes exhibit only a rather small discontinuity [see
\citefig{fig3}(a)]. Therefore,
in the following, we
concentrate on  neutron shell gaps obtained
from the nuclear masses around
$^{40}_{20}$Ca, $^{48}_{28}$Ca, $^{56}_{28}$Ni, $^{90}_{50}$Zr,
$^{132}_{~82}$Sn, and $^{208}_{126}$Pb.
The direct information on the
isotopic dependence of the {\shellgap} sizes
is even more restrained. Only a comparison between the {\shellgap} values for
$^{56}$Ni and $^{48}$Ca
suggests that
the size of
the $N$=28 shell gap decreases with increasing neutron excess.

In \citefig{fig3}(b), we compare, with experimental data, the results
of global mass calculations
obtained in
the finite-range droplet model (FRDM) of Ref.{\spc}\cite{[Mol93]}.
This is a microscopic-macroscopic  model which is known to give a very
good fit to experimental nuclear masses.
(It is worth noting that the FRDM explicitly considers
deformation effects.) Consequently, the
two-neutron separation energies
are also well reproduced in this model.
 Only in some
specific regions of the periodic chart are
some discrepancies present. For example, a sudden
decrease in  $S_{2n}$ around $^{96}$Zr is not reproduced and also
the trend  predicted for light zirconium isotopes is slightly off.
As far as the jumps of $S_{2n}$ at magic numbers, the FRDM
reproduces
the six data points fairly well. In particular,
the best agreement is obtained for  $^{90}$Zr;
the largest discrepancy is seen in
$^{48}$Ca.

At this point,  one should note that any kind of correlation,
absent in a magic nucleus but appearing   in its two even neighbors,
would lead to a decreased size of the shell gap (\ref{e3}). Indeed,
in such a situation, the
binding energy of the magic nucleus is
expected to be well described by
an uncorrelated state, while the correlations
in the neighboring nuclei would   lead  to a reduced  value of
$S_{2n}^N$$-$$S_{2n}^{N+2}$.
Consequently, if correlations
(e.g., fluctuations due to pairing vibrations, shape oscillations, etc.)
are missing in the model description,
 the differences $S_{2n}^N$$-$$S_{2n}^{N+2}$ can easily be
overestimated.
Of course, this kind of reasoning tacitly assumes that the
 effects on {\shellgap} sizes originating from the pure mean field
are correctly described
which certainly need not to be valid.

Predictions of the FRDM concerning the isotopic dependence of {\shellgap}
sizes are very clear. With increasing neutron excess, the gap at
$N$=28 decreases (too much as compared with experiment), while the gaps at
$N$=50 and 82 stay practically the same.
In practical terms, in the microscopic-macroscopic
methods the {\shellgap} sizes are
the direct consequences of simple parametrizations of
the isotopic dependence of one-body average potential. Such dependences
probably cannot be inferred, with sufficiently high precision,
from fits to known nuclei, because these nuclei form isotopic chains
which are not long enough.
{}From this point of view,
far-reaching extrapolations, based on two-body density-dependent
effective interactions,
have a better chance of success.
Moreover, the results of microscopic-macroscopic methods are
plagued with numerous uncertainties
when they are applied to nuclei far from stability \cite{[Naz94]}.

In \citefig{fig3}(c),  we present results obtained by
 the spherical HFB+SkP calculations. Here,
the overall agreement with data is  similar to that obtained in the
{}FRDM. Apart from slightly too high values of $S_{2p}$
obtained before the
$N$=126 shell closure in the lead isotopes, and small deviations
for some zirconium isotopes (similar as in the FRDM),
the isotopic dependence of the
two-neutron separation energies is reproduced fairly well.
The predicted size of the magic gap
is slightly too large in $^{208}$Pb, slightly too small in
$^{90}$Zr and  $^{132}$Sn, and seriously underestimated in the light
magic nuclei. Predictions concerning the isotopic dependence
of the {\shellgap} sizes are markedly different
than those of the FRDM.
With increasing neutron excess,  the shell gaps at $N$=50 and 82
decrease substantially. This effect has already been observed
in Refs.{\spc}\cite{[Smo93],[Dob94]} and is discussed in
\citesec{sec4}.

After adjusting the strength of the contact force (\ref{e1})
to the odd-even mass staggering (\citesec{sec2b}),
we are now able to perform the HFB calculations with other standard
Skyrme parametrizations. The results for the SIII$^\delta$
and SkM$^\delta$ forces  are
presented in \citefigs{fig4}(a) and \ref{fig4}(b), respectively. They are
compared with the results of the SkP$^\delta$ force
shown in \citefig{fig4}(c).
One should note that the results obtained with the
SkP and SkP$^\delta$ forces are very similar,
in spite of their very different properties in the pairing channel.

As seen in \citefig{fig4}, neither for SIII$^\delta$ nor for
SkM$^\delta$ is the agreement with data
satisfactory.
In particular, the {\shellgap} sizes at $N$=50, 82, and 126
are strongly overestimated, and
 the values and the slopes of $S_{2n}$
are, in most cases, incorrect.
At this point,
one should note that a systematic few-MeV discrepancy in two-neutron
separation energies leads to an integrated many-MeV discrepancy
in total binding energies, cf.{\spc}results for lead isotopes presented in
Ref.{\spc}\cite{[Taj93b]}.

The results shown in \citefig{fig4}
illustrate a very strong dependence of the
two-neutron separation energies on the force parameters.
Although older forces, such as SIII and SkM*,
can perform well in certain regions of $Z$ and $N$, they
do not give a satisfactory global reproduction
of the data. On the other hand,
a fairly good global agreement obtained with  the SkP suggests that the
improvement is possible, while a still better parametrization
would be welcome (see Refs.{\spc}\cite{[Mey94],[Rei94]}).
Of course, forces which fail in reproducing the
behavior with ($N-Z$) in known nuclei have little chance to perform
better
when going far from stability. For example, the almost constant
values of $S_{2n}$ obtained in the very heavy tin and lead isotopes
with the SIII$^\delta$ and  SkM$^\delta$
interactions, followed by strong
shell effects at $N$=126 and $N$=184, do not seem
very reliable.
A  detailed analysis of the force-dependence
of results  may give us
valuable
information on the relative importance of various force parameters.
In the next Section,  such an analysis is performed
for  the sizes of
magic gaps in known and exotic nuclei.

\section{Magic shell gaps far from stability}
\label{sec4}

In the present Section, the $Z$-dependence  of sizes of neutron
shell gaps is analyzed. The {\shellgap} sizes are
defined as in \citeq{e3};  six neutron
magic numbers from $N$=20 to 126 are considered. In the
$Z$-representation,
neutron-rich nuclei appear at small proton numbers and the proton drip
line is approached at  large proton numbers.

\subsection{The role of deformation}
\label{sec4a}

Experimental values of
$S_{2n}^N$$-$$S_{2n}^{N+2}$
as functions of $Z$
are shown in \citefig{fig5}(a).
A conspicuous feature present
in the data is the appearance  of well-defined maxima
at magic numbers.
This fact can be associated with the effect of deformation.
Namely, the {\shellgap} sizes defined in \citeq{e3} depend
on the masses of even isotopes adjacent to a neutron-magic
system. In nuclei which are not proton-magic,
both neutron and proton shells are open. In such cases,
deformation effects have to be considered.
Therefore, the isotopic dependencies seen in
\citefig{fig5}(a) illustrate
not only  the ``pure'' isotopic dependence of the shell effects
but also
the magnitude  of deformation-like
correlations.
Of course, experimentally, one cannot easily disentangle the deformation
and isotopic effects, and the complete theoretical description should
take them into account simultaneously. On the other hand, extrapolations
to nuclei far from stability are difficult enough, and one would wish
to consider one problem at a time. Since, in the present study, we
consider spherical shapes only,
 we must disregard the isotopic
dependencies present in the data. Consequently,
we concentrate mainly on  the six cases of
doubly magic nuclei discussed in \citesec{sec3}. In \citefigs{fig5}(b)
and \ref{fig5}(c), and also in the figures discussed below,
these six values are denoted by full circles and, in addition, two
points corresponding to the $N$=28 systems are connected by a thick line.

In spite of the fact that
 the FRDM results account for  the deformation effects, this model
does not reproduce the experimental
differences $S_{2n}^N$$-$$S_{2n}^{N+2}$
in an entirely satisfactory way. The maxima at $N$=20 and $N$=$Z$=28
are not high enough and the maximum at $^{48}$Ca is absent. The
isotopic trend predicted for the light $N$=50 isotones is correct, but the
proton-number dependence in the
heavy $N$=50 isotones, and in the $N$=82 and 126 isotones, is visibly
too weak. One should, of course, bear in mind that we are now looking
at very fine details of mass predictions.

The HFB+SkP results presented in \citefig{fig5}(c)
show very clearly that sizes of spherical shell gaps
strongly depend on the neutron excess. For all neutron magic numbers,
the gap sizes decrease when approaching the neutron drip line; i.e.,
when the proton number decreases for a fixed value of $N$.
(Again, since these results are restricted to
spherical shapes, one cannot expect
to reproduce the experimental
proton-number dependence seen in \citefig{fig5}(a).)
      In the automated calculations we have performed for long chains
          of isotopes or isotones,
in some cases, a poor convergence was obtained,
                due to crossings between  single-particle or
                quasiparticle levels.
Since, in this work, we discuss only general trends, we do not try
                to cure  these cases. (This would require
the full  configuration
                blocking.) In \protect\citefigs{fig5}-\protect\ref{fig8}
                the converged points are connected by straight lines.

\subsection{Hartree-Fock results}
\label{sec4b}

We are now in a position to discuss possible origins of the isospin
dependence of the {\shellgap} sizes. First of all,
by using the Hartree-Fock (HF)
method in which the pairing correlations are neglected,
we may study the
role of the interactions in the p-h channel alone. In \citefig{fig6}
we present the HF
results for three  Skyrme forces,
namely, SIII, SkM*, and SkP.

Let us first discuss the force-independent, generic features
of the predicted {\shellgap} sizes.
{}For all the forces considered,
 the sizes of the $N$=20 to $N$=82 shell
gaps decrease with decreasing proton number. The effect is
strongest for $N$=28,
and then weakens for $N$=50 and $N$=82. For $N$=126, the gap size
is predicted to be almost constant  as a function of $Z$,
and even a reversed trend is obtained in  the HF+SIII model.
In all cases,
a decrease of the $N$=28 shell gap between $^{56}$Ni and
$^{48}$Ca is predicted.

Systematically, the magnitude of the  shell effect is too strong for the
SIII and SkM* interactions. The sizes of the $N$=50 and $N$=82
shell gaps are overestimated by as much as a factor of two.
A part of this discrepancy can be attributed to  missing correlations.
As discussed
above, they always decrease the {\shellgap} sizes; the influence of pairing
correlations is discussed  in \citesec{sec4c}.
In  the HF+SkP model, the size of the
$N$=28 shell gap is too small.
  One may therefore
anticipate
that it will be much too small after the pairing correlations are
taken into account.

Overall magnitudes of {\shellgap} sizes can be clearly correlated with
different values of the effective mass corresponding to the three studied
Skyrme forces.
Indeed, the single-particle level density $g$ at the Fermi level can be
approximated by \cite{[Gra79],[Jen86]}
\be
g = {3\over 4}N_t\frac{2m^*}{(\hbar k_F)^2},
\ee
where $N_t$ is the particle number, $k_F$ is the Fermi momentum, and
$m^*$ is the effective mass.
{}For the SIII, SkM*, and SkP, the effective
masses are   $m^*/m$=0.76, 0.79, and 1.0,
respectively. Therefore, the density of single-particle states increases
when going from SIII to SkP, and hence the sizes of shell gaps decrease
accordingly.

Interesting conclusions can be drawn when comparing sizes of the
$N$=20 and $N$=28 shell gaps. In experiment, the $N$=20 gap  is clearly
larger (in $^{40}$Ca) than the $N$=28 gap
(in $^{48}$Ca), which is the feature reproduced
by the HF+SkP model. This means that the neutron
f$_{7/2}$ orbital is closer to the upper major shell than to the lower one.
On the other hand, the SIII and SkM* forces
give similar sizes for both shell gaps,
which may suggest that these forces place this orbital too low and have
therefore too large a strength
of the spin-orbit interaction (cf. discussion in Ref.{\spc}\cite{[Rei94]}).

\subsection{Hartree-Fock-Bogoliubov results}
\label{sec4c}

By comparing results presented in
\citefigs{fig5}(c) and \ref{fig6}(c), one may see that the
pairing correlations dramatically influence the {\shellgap} sizes.
In light nuclei, the gaps are reduced by about a factor of two,
which gives substantial disagreement with experiment.
One may attribute this effect to two factors: Firstly, the sizes
of the $N$=20 and $N$=28 gaps  obtained in  the HF method are  too
small. This may suggest that in light nuclei the effective interaction
should have an
effective mass smaller than unity; this is the case  for the SIII and SkM*
forces.
Since both predicted shell gaps at
$N$=20 and $N$=28 seem to be too small at the same time,
 the discrepancy with experiment
should not be attributed to an incorrect placement of the f$_{7/2}$
shell, i.e., to an incorrect spin-orbit strength.
Secondly,
the pairing matrix elements of the SkP force are probably too strong
in light nuclei.
This has already been discussed in \citesec{sec2a} in connection with
the nonvanishing pairing correlations in the neutron-magic nuclei.
Of course, the predicted strong pairing correlations can also
result from reduced  shell gaps, and both explanations
strongly depend on one another.

In heavier nuclei, pairing correlations bring the  {\shellgap} sizes,
predicted by the HF+SkP model,
very close to experimental data. They also induce a dramatic quenching
of shell effects in neutron-rich nuclei. For example,
in the HF+SkP model, the  $N$=82 shell
gap decreases from about 9~MeV  to 5~MeV, with $Z$ going down from
86 to 40. In the HFB+SkP method, the $N$=82  gap decreases from 7~MeV to
almost zero. As a result, there is almost no influence
of the $N$=82 gap on the
position of the spherical neutron drip line  \cite{[Smo93]}.

{}Figures \ref{fig7} and \ref{fig8} display
the HFB results for the
{\shellgap} sizes obtained with the density-independent (\ref{e1})
and density-dependent (\ref{e2}) contact forces, respectively.
The results are presented for the three Skyrme interactions discussed
above.
It can be seen that for the interactions
SIII and SkM*
which have low effective mass,
 pairing correlations are not sufficiently strong to
bring the shell gaps in heavy nuclei down to the experimental values.
In light nuclei, the sizes of the $N$=20 and $N$=28 shell gaps are again
almost
equal, and therefore either the former (for SIII$^\delta$) or the
latter (for SkM$^\delta$) values agree with experiment, but a consistent
description cannot be obtained.

In general, the {\shellgap} sizes obtained with  the
density-dependent pairing interaction are lower than those obtained
with  the
pure contact force. The differences between  different p-p
interactions are significant. For example, the
$N$=82 shell gaps
predicted in the SkM$^\delta$ calculations
are always greater than 5~MeV, while those obtained in the
SkM$^{\delta\rho}$ version exhibit a pronounced  quenching when
approaching the neutron drip line.

\section{Conclusions}
\label{sec5}

The results of self-consistent HF and HFB calculations with
several Skyrme-type interactions
demonstrate that the magnitude of shell effects at magic numbers
strongly depends on force parameters. In the p-h channel, the
main factor is the effective mass which determines the
single-particle  level
density around the Fermi level. In the pairing channel, the
density-dependence of a p-p interaction plays a very important role.
Both effects combined, i.e., the increase in $m^*/m$ and
the presence of density-dependent pairing, lead to the
shell-quenching effects in nuclei far from the beta-stability line.
This result seems to be  consistent with the recent analysis
of solar-system r-process abundance distributions
\cite{[Kra94]}.

Our brief analysis indicates
 that a lot of systematic work still needs to be done
to obtain a ``universal"  Skyrme-type interaction. For instance,
the density dependence of the effective interaction
invites a much
deeper analysis. (In principle,
the force should depend on both the isoscalar density,
$\rho_n+\rho_p$, and the isovector density, $\rho_n-\rho_p$
\cite{[Dab77],[Reg88]};
the latter might be important in the  description of exotic nuclei
with a significant neutron excess.) Also, the interplay
between deformation and pairing in open-shell drip-line nuclei
is still awaiting explanation.

\section*{Acknowledgements}

This research was supported in part by the
Polish State Committee for Scientific Research under Contract
No. 20450~91~01.
Oak Ridge National
Laboratory is managed for the U.S. Department of Energy by Martin
 Marietta Energy Systems, Inc. under Contract No.
DE-AC05--84OR21400.
The Joint Institute for Heavy Ion
 Research has as member institutions the University of Tennessee,
Vanderbilt University, and the Oak Ridge National Laboratory; it
is supported by the members and by the Department of Energy
through Contract No. DE-FG05-87ER40361 with the University
of Tennessee.  Theoretical nuclear physics research
at the University of Tennessee
 is supported by the U.S. Department of
Energy through Contract No. DE-FG05-93ER40770.

%\bibliography{wn}
%\bibliographystyle{unsrt}

\clearpage
\newpage

\begin{table}
\caption[T2]{Experimental and calculated neutron odd-even mass
difference
            $\delta_{\rm N}^{\rm exp}$  and
             $\delta_{\rm N}$ for selected proton-magic nuclei.
Experimental pairing gaps $\Delta_{\rm N}^{\rm exp}$
(determined as in \protect\cite{[Mol92b]}) and calculated
             neutron and proton average pairing gaps
             $\tilde{\Delta}_{\rm{N}}$ and
             $\tilde{\Delta}_{\rm{P}}$ are also given.
\label{tab2}}

\vspace{1cm}
\begin{center}
\begin{tabular}{cc|ccccc}
\hline
 $N$ & $Z$ & $\delta_{\rm N}^{\rm exp}$ &$\delta_{\rm N}$ &
                                         $\Delta_{\rm N}^{\rm exp}$ &
                                         $\tilde{\Delta}_{\rm N}$ &
                                         $\tilde{\Delta}_{\rm P}$  \\
\hline
  23  &  20  &  1.600  &  1.301  &  1.708  &  1.584  &  0.000  \\
  31  &  28  &  1.194  &  0.234  &  1.446  &  0.720  &  0.000  \\
  53  &  40  &  0.743  &  0.448  &  0.829  &  0.712  &  0.793  \\
  69  &  50  &  1.312  &  1.293  &  1.378  &  1.256  &  0.000  \\
 121  &  82  &  0.736  &  0.634  &  0.801  &  0.737  &  0.000  \\
\hline
\end{tabular}
\end{center}

\end{table}

%\clearpage

\begin{table}
\caption[T3]{Same as in \protect\citetab{tab2}, except for protons replacing
neutrons and {\em vice versa}.
\label{tab3}}

\vspace{1cm}
\begin{center}
\begin{tabular}{cc|ccccc}
\hline
 $N$ & $Z$ & $\delta_{\rm P}^{\rm exp}$ &$\delta_{\rm P}$ &
                                         $\Delta_{\rm P}^{\rm exp}$ &
                                         $\tilde{\Delta}_{\rm P}$ &
                                         $\tilde{\Delta}_{\rm N}$  \\
\hline
  20  &  17  &  0.929  &  0.496  &  1.535  &  1.191  &  1.202  \\
  28  &  25  &  1.147  &  0.709  &  1.540  &  1.139  &  0.605  \\
  40  &  31  &  0.938  &  0.303  &  1.385  &  0.900  &  1.319  \\
  50  &  37  &  0.997  &  0.514  &  1.356  &  0.976  &  0.000  \\
  82  &  59  &  0.998  &  0.664  &  1.229  &  0.931  &  0.000  \\
 126  &  85  &  0.661  &  0.530  &  0.846  &  0.552  &  0.000  \\
\hline
\end{tabular}
\end{center}

\end{table}

%\clearpage

\begin{table}
\caption[T1]{Parameters of pairing forces used in the present study.
             Columns denoted by $\delta$ and $\delta\rho$ give
             parameters of forces (\protect\ref{e1}) and
             (\protect\ref{e2}), respectively. In the latter
             case, parameters $V_0$ and $\gamma$ are identical
             to those in the p-h channel. Different sets of
             parameters are fitted to different forces used in
             the p-h channel, as indicated in the first column.
\label{tab1}}

\vspace{1cm}
\begin{center}\begin{tabular}{c||c|ccc}
\hline
{}Force in the &$\delta$&         \multicolumn{3}{c}{$\delta\rho$}          \\
                                                                 \cline{2-5}
p-h channel  & $V_0$~(MeV~fm$^3$)
             & $V_0$$\equiv$$t_0$~(MeV~fm$^3$)
             & $V_3$~(MeV~fm$^{3+\gamma}$)
             & $\gamma$$\equiv$$\alpha$                                     \\
                                                                 \hline
SIII         &  -160  &     -1128.75      &  50715  &   1                   \\
SkM*         &  -173  &     -2645         &  21591  &   1/6                 \\
SkP          &  -185  &     -2931.696     &  24175  &   1/6                 \\
                                                                 \hline
\end{tabular}\end{center}
\end{table}

\clearpage
\newpage

\begin{center}{Figure captions}\end{center}

\begin{figure}[htb]
\caption[F1]{Average neutron pairing gaps, $\tDeltan$, for
             proton-magic nuclei listed in \protect\citetab{tab2}.
             Odd-even staggering parameter $\delta_{\rm N}^{\rm
             exp}$ is plotted as the experimental value. All
             forces were adjusted so as to give identical
             results for $^{119}_{~69}$Sn$_{50}$.
\label{fig1}}
\end{figure}

\begin{figure}[htb]
\caption[F2]{Average proton pairing gaps, $\tDeltap$, for
             neutron-magic nuclei listed in \protect\citetab{tab3}.
             Odd-even staggering parameter $\delta_{\rm P}^{\rm
             exp}$ is plotted as the experimental value.
\label{fig2}}
\end{figure}

\begin{figure}[htb]
\caption[F3]{Two-neutron separation energies, $S_{2n}$, for
             proton-magic isotopes. In (a), experimental (full circles)
             and ``from systematics'' \protect\cite{[Aud93]} (open circles)
             values are shown. In (b) and (c)
compared with the experimental data are
the results of
             the FRDM \protect\cite{[Mol93]} and of the HFB+SkP model,
             respectively.
\label{fig3}}
\end{figure}

\begin{figure}[htb]
\caption[F4]{Two-neutron separation energies, $S_{2n}$, for
             proton-magic isotopes. The HFB results with
             the SIII$^\delta$ (a), SkM$^\delta$ (b), and
             SkP$^\delta$ (c) parametrizations (see \protect\citesec{sec2b})
             are compared with the experimental data.
\label{fig4}}
\end{figure}

\begin{figure}[htb]
\caption[F5]{Sizes of neutron shell gaps as functions of the number of
             protons. Experimental data (a) are compared with the
             results of
             the FRDM \protect\cite{[Mol93]} (b) and of the HFB+SkP method
(c).
             Deformation effects are included in the FRDM while only
             spherical shapes are considered in HFB+SkP. In (b) and (c),
             experimental data are shown only for magic nuclides and
             the thick line  connects experimental
points for two magic $N$=28 systems.
\label{fig5}}
\end{figure}

\begin{figure}[htb]
\caption[F6]{Sizes of neutron shell gaps obtained within the HF approximation
             for three parametrizations of the Skyrme force:
             SIII (a), SkM* (b), and SkP (c).
             Experimental data are shown only for magic nuclides and
             the thick line connects  two experimental $N$=28 points.
\label{fig6}}
\end{figure}

\begin{figure}[htb]
\caption[F7]{Sizes of neutron shell gaps obtained within the HFB approximation
             for three parametrizations of the Skyrme force:
             SIII$^\delta$ (a), SkM$^\delta$ (b), and SkP$^\delta$ (c).
             Contact pairing force (\protect\ref{e1}) is used. The
             same experimental data as in \protect\citefig{fig6} are shown.
\label{fig7}}
\end{figure}

\begin{figure}[htb]
\caption[F8]{Sizes of neutron shell gaps obtained within the HFB approximation
             for three parametrizations of the Skyrme force:
             SIII$^{\delta\rho}$ (a), SkM$^{\delta\rho}$ (b),
             and SkP$^{\delta\rho}$ (c).
             Density-dependent contact pairing force (\protect\ref{e2})
             is used. The
             same experimental data as in \protect\citefig{fig6} are shown.

             \vspace{15cm}
\label{fig8}}
\end{figure}

\disregard{
\clearpage
\newpage

\centerline{\bf\Large Note to the Editors}

\vspace{1cm}
\centerline{
Please print Figures 3 and 4 using the full width of the Journal page.}
}

\end{document}